\documentclass[prc,preprint]{revtex4}
\usepackage{graphics}

\newcommand{\be}{\begin{equation}}
\newcommand{\ee}{\end{equation}}

\begin{document}

\title{{\bf
Investigation of the field-induced ferromagnetic phase transition
in spin polarized neutron matter: a lowest order constrained
variational approach}}

\author{{\bf G.H. Bordbar $^{1,2}$ \footnote{Corresponding author. E-mail:
bordbar@physics.susc.ac.ir}}, {\bf Z. Rezaei$^{1}$} and {\bf Afshin
Montakhab$^{1}$}}
 \affiliation{ $^1$Department of Physics,
Shiraz University,
Shiraz 71454, Iran\footnote{Permanent address},\\
 and\\
$^2$Research Institute for Astronomy and Astrophysics of Maragha,\\
P.O. Box 55134-441, Maragha, Iran}


\begin{abstract}
In this paper, the lowest order constrained variational (LOCV)
method has been used to investigate the magnetic properties of spin
polarized neutron matter in the presence of strong magnetic field at
zero temperature employing $AV_{18}$  potential. Our results
indicate that a ferromagnetic phase transition is induced by a
strong magnetic field with strength greater than $10^{18}\ G$,
leading to a partial spin polarization of the neutron matter.
It is also shown that the equation of state  of neutron matter in
the presence of magnetic field is stiffer than the case in absence
of magnetic field.
\end{abstract}
\pacs{21.65.-f, 26.60.-c, 64.70.-p}

\maketitle

\section{INTRODUCTION}
The magnetic field of neutron stars  most probably originates from
the compression of magnetic flux inherited from the progenitor star
\cite{Reisen}. Using this point of view, Woltjer has predicted a
magnetic field strength of order $10^{15}\ G$ for the neutron stars
\cite{Woltjer}. The field can be distorted or amplified by some
mixture of  convection, differential rotation and magnetic
instabilities \cite{Tayler,Spruit}. The relative importance of these
ingredients depend on the initial field strength and rotation rate
of the star. For both convection and differential rotation, the
field and its supporting currents are not likely to be confined to
the solid crust of the star, but distributed in most of the stellar
interior which is mostly a fluid mixture of neutrons, protons,
electrons, and other more exotic particles \cite{Reisen}. Thompson
et al. \cite{Thompson} argued that the newborn neutron stars
probably combine vigorous convection and differential rotation
making it likely that a dynamo process might operate in them. They
expected fields up to $10^{15}-10^{16}\ G$ in neutron stars with few
millisecond initial periods. On the other hand, according to the
scalar virial theorem which is based on Newtonian gravity, the
magnetic field strength is allowed by values up to $10^{18}\ G$ in
the interior of a magnetar \cite{Lai}. However, general relativity
predicts the allowed maximum value of neutron star magnetic field to
be about $10^{18}-10^{20}\ G$ \cite{shap}.  By comparing with the
observational data, Yuan et al.  \cite{Yuan} obtained a magnetic
field strength of order $10^{19}\ G$ for the neutron stars.

The strong magnetic field could have important influences on the
interior matter of a neutron star. Many works have dealt with study
of the magnetic properties and  equation of state of the neutron
star matter \cite{zhang,Brod0,SUH,Chen,Yue6,
Brod2,Chakra7,Isayev,Isayev1,Garcia8,Garcia9,Garcia10} and quark
star matter \cite{ANAND,Ghosh,Chakra6,Gupta,Bandyo,bord-pey} in the
presence of strong magnetic fields. Some authors have considered the
influence of strong magnetic fields on the neutron star matter
within the mean field approximation \cite{zhang,Chen}. Yuan et al.
\cite{zhang} using the nonlinear $\sigma-\omega$ model, showed that
the equation of state of neutron star matter becomes softer as the
magnetic field increases. Also, Broderick et al. \cite{Brod0}
employing a field theoretical approach in which the baryons interact
via the exchange of $\sigma-\omega-\rho$ mesons, observed that the
softening of the equation of state caused by Landau quantization is
overwhelmed by stiffening due to the incorporation of the anomalous
magnetic moments of the nucleons. It has been shown that the strong
magnetic field shifts $\beta$-equilibrium and increases the proton
fraction in the neutron star matter \cite{Brod0,SUH,Chen}.
Yue et al. \cite{Yue6} have studied the neutron star matter in the
presence of strong magnetic field using the quark-meson coupling
(QMC) model. Their results indicate that the Landau quantization of
charged particles causes a softening in the equation of state,
whereas the inclusion of nucleon anomalous magnetic moments lead to
a stiffer equation of state. The effects of the magnetic field on
the neutron star structure, through its influence on the metric has
been studied by Cardall et al. \cite{Cardall}. Their results show
that the maximum mass, in a static configuration for neutron star
with magnetic field, is larger than the maximum mass obtained by
uniform rotation.
Through a field theoretical approach (at the mean field level) in
which the baryons interact via the exchange of $\sigma-\omega-\rho$
mesons, Broderick et al. \cite{Brod2} have considered the effects of
magnetic field on the equation of state of dense baryonic matter in
which hyperons are present. They found that when the hyperons
appear, the pressure becomes smaller than the case of pure nucleonic
matter for all fields. Within a relativistic Hartree approach in the
linear $\sigma-\omega-\rho$ model, the effects of magnetic field on
cold symmetric nuclear matter and the nuclear matter in
$\beta$-equilibrium have been investigated by Chakrabarty et al.
\cite{Chakra7}. Their results suggest that the neutron star mass is
practically insensitive to the effects of the magnetic fields,
whereas the radius decreases in intense fields .

In some studies, the neutron star matter was approximated by a
pure neutron matter. Isayev et al. \cite{Isayev} considered the
neutron matter in a strong magnetic field with the Skyrme
effective interaction and analyzed the resultant self-consistent
equations. They found that the thermodynamically stable branch
extends from the very low densities to the high density region
where the spin polarization parameter is saturated, and neutrons
become totally spin polarized.
Perez-Garcia et al. \cite{Garcia8,Garcia9,Garcia10} studied the
effects of a strong magnetic field on the pure neutron matter with
effective nuclear forces within the framework of the nonrelativistic
Hartree-Fock approximation. They showed that in the Skyrme model
there is a ferromagnetic phase transition at $\rho \sim
4\rho_{0}$($\rho_{0}=0.16 fm^{-3}$ is the nuclear saturation
density), whereas it is forbidden in the $D1P$ model \cite{Garcia8}.
Beside these, they found that the neutrino opacity of magnetized
matter decreases compared to the nonmagnetized case for the magnetic
field greater than $10^{17}\ G$ \cite{Garcia9}.
However, more realistically, for the problem of the neutron star
matter in astrophysics context, it is necessary to consider the
finite temperature \cite{Isayev1,Garcia8,Chakra6,Gupta} and finite
proton fraction effects
\cite{Brod2,Chakra7,zhang,Brod0,SUH,Chen,Yue6}. Isayev et al.
\cite{Isayev1} have shown that the influence of finite temperatures
on spin polarization remains moderate in the Skyrme model, at least
up to temperatures relevant for protoneutron stars. It has been also
shown that for $SLy4$ effective interaction, even small admixture of
protons to neutron matter leads to a considerable shift of the
critical density of the spin instability to lower values. For $SkI5$
force, however, a small admixture of protons to neutron matter does
not considerably change the critical density of the spin instability
and increases its value \cite{Isayev2}.

In our previous works, we have studied the spin polarized neutron
matter \cite{Bordbar75}, symmetric nuclear matter \cite{Bordbar76},
asymmetric nuclear matter \cite{Bordbar77}, and neutron star matter
\cite{Bordbar77} at zero temperature using LOCV method with the
realistic strong interaction in the absence of magnetic field. We
have also investigated the thermodynamic properties of the spin
polarized neutron matter \cite{Bordbar78}, symmetric nuclear matter
\cite{Bordbar80}, and asymmetric nuclear matter \cite{Bordbar81} at
finite temperature with no magnetic field. In the above
calculations, our results do not show any spontaneous ferromagnetic
phase transition for these systems. In the present work, we study
the magnetic properties of spin polarized neutron matter at zero
temperature in the presence of the strong magnetic field using LOCV
technique employing $AV_{18}$ potential.

\section{  LOCV formalism for spin polarized neutron matter  }
We consider a pure homogeneous spin polarized neutron matter
composed of the spin-up $(+)$ and spin-down $(-)$ neutrons. We
denote the number densities of spin-up and spin-down neutrons by
$\rho^{(+)}$ and $\rho^{(-)}$, respectively. We introduce the spin
polarization parameter ($\delta$) by
\begin{eqnarray}
     \delta=\frac{\rho^{(+)}-\rho^{(-)}}{\rho},
 \end{eqnarray}
where $-1\leq\delta\leq1$, and $\rho=\rho^{(+)}+\rho^{(-)}$ is the
total density of  system.

In order to calculate the energy of this system, we use LOCV method
as follows: we consider a trial many-body wave function of the form
\begin{eqnarray}
     \psi=F\phi,
 \end{eqnarray}
where $\phi$   is the uncorrelated ground-state wave function of
$N$ independent neutrons, and $F$ is a proper $N$-body correlation
function. Using Jastrow approximation \cite{Jastrow}, $F$ can be
replaced by
\begin{eqnarray}
    F=S\prod _{i>j}f(ij),
 \end{eqnarray}
 where $S$ is a symmetrizing operator. We consider a cluster expansion of the
 energy functional up to the two-body term,
 \begin{eqnarray}\label{tener}
           E([f])=\frac{1}{N}\frac{\langle\psi|H|\psi\rangle}
           {\langle\psi|\psi\rangle}=E _{1}+E _{2}\cdot
 \end{eqnarray}
 Now, we calculate the energy per particle up to the two-body term for two
 cases in the absence and presence of the magnetic field in two separate sections.
\subsection{Energy calculation for  the spin polarized neutron matter in the absence of magnetic field}
The one-body term $E_{1}$ for spin polarized neutron matter in the
absence of magnetic field $(B=0)$ is given by
 \begin{eqnarray}
 \label{oneterm}
E_{1}^{(B=0)}=\sum_{i=+,-}\frac{3}{5}\frac{\hbar^{2}k_{F}^{(i)^{2}}}{2m}\frac{\rho^{(i)}}{\rho},
\end{eqnarray}
where $k_{F}^{(i)}=(6\pi^{2}\rho^{(i)})^{\frac{1}{3}}$ is the
Fermi momentum of a neutron with spin projection $i$.

The two-body energy $E_{2}$ is
 \begin{eqnarray}
    E_{2}^{(B=0)}&=&\frac{1}{2N}\sum_{ij} \langle ij\left| \nu(12)\right|
    ij-ji\rangle,
 \end{eqnarray}
where
$$\nu(12)=-\frac{\hbar^{2}}{2m}[f(12),[\nabla
_{12}^{2},f(12)]]+f(12)V(12)f(12).$$ In the above equation, $f(12)$
and $V(12)$ are the two-body correlation function and nuclear
potential, respectively. In our calculations, we employ the
$AV_{18}$ two-body potential \cite{Wiringa},
\begin{eqnarray}
V(12)&=&\sum^{18}_{p=1}V^{(p)}(r_{12})O^{(p)}_{12}.
\end{eqnarray}
where
\begin{eqnarray}\label{operat18}
O^{(p=1-18)}_{12}&=&1,\sigma_{1}.\sigma_{2},\tau_{1}.\tau_{2},
(\sigma_{1}.\sigma_{2})(\tau_{1}.\tau_{2})
,S_{12},S_{12}(\tau_{1}.\tau_{2}),\nonumber\\&&\textbf{L}.\textbf{S},
\textbf{L}.\textbf{S}(\tau_{1}.\tau_{2}),\textbf{L}^{2},
\textbf{L}^{2}(\sigma_{1}.\sigma_{2}),\textbf{L}^{2}(\tau_{1}.\tau_{2}),
\textbf{L}^{2}(\sigma_{1}.\sigma_{2})(\tau_{1}.\tau_{2}),
\nonumber\\&&(\textbf{L}.\textbf{S})^{2},(\textbf{L}.\textbf{S})^{2}
(\tau_{1}.\tau_{2}),\textbf{T}_{12},(\sigma_{1}.\sigma_{2})\textbf{T}_{12},
S_{12}\textbf{T}_{12},(\tau_{z1}+\tau_{z2}).
\end{eqnarray}
In the above equation,
$$S_{12}=[3(\sigma_{1}.\hat{r})(\sigma_{2}.\hat{r})-\sigma_{1}.\sigma_{2}]$$
is the tensor operator and
$$\textbf{T}_{12}=[3(\tau_{1}.\hat{r})(\tau_{2}.\hat{r})-\tau_{1}.\tau_{2}]$$
is the isotensor operator. The above $18$ components of the
$AV_{18}$ two-body potential are denoted by the labels $c$,
$\sigma$, $\tau$, $\sigma\tau$, $t$, $t\tau$, $ls$, $ls\tau$,
$l2$, $l2\sigma$, $l2\tau$, $l2\sigma\tau$, $ls2$, $ls2\tau$, $T$,
$\sigma T$, $tT$, and $\tau z$, respectively \cite{Wiringa}. In
the LOCV formalism, the two-body correlation function $f(12)$ is
considered as follows \cite{Owen},
\begin{eqnarray}
f(12)&=&\sum^3_{k=1}f^{(k)}(r_{12})P^{(k)}_{12},
\end{eqnarray}
where
\begin{eqnarray}
P^{(k=1-3)}_{12}&=&(\frac{1}{4}-\frac{1}{4}O_{12}^{(2)}),\
(\frac{1}{2}+\frac{1}{6}O_{12}^{(2)}+\frac{1}{6}O_{12}^{(5)}),\
(\frac{1}{4}+\frac{1}{12}O_{12}^{(2)}-\frac{1}{6}O_{12}^{(5)}).
\end{eqnarray}
The operators $O_{12}^{(2)}$ and $O_{12}^{(5)}$ are given in Eq.
(\ref{operat18}). Using the above two-body correlation function and
potential, after doing some algebra, we find the following equation
for the two-body energy:
\begin{eqnarray}\label{ener2}
    E_{2}^{(B=0)} &=& \frac{2}{\pi ^{4}\rho }\left( \frac{\hbar^{2}}{2m}\right)
    \sum_{JLSS_{z}}\frac{(2J+1)}{2(2S+1)}[1-(-1)^{L+S+1}]\left| \left\langle
\frac{1}{2}\sigma _{z1}\frac{1}{2}\sigma _{z2}\mid
SS_{z}\right\rangle \right| ^{2}  \nonumber
\\&& \int dr\left\{\left [{f_{\alpha
}^{(1)^{^{\prime }}}}^{2}{a_{\alpha
}^{(1)}}^{2}(k_{f}r)\right.\right.\left.\left.
+\frac{2m}{\hbar^{2}}(\{V_{c}-3V_{\sigma } +V_{\tau }-3V_{\sigma
\tau }+2(V_{T}-3V_{\sigma T }) -2V_{\tau z}\}{a_{\alpha
}^{(1)}}^{2}(k_{f}r)\right.\right. \nonumber \\&&\left.\left.
+[V_{l2}-3V_{l2\sigma } +V_{l2\tau }-3V_{l2\sigma \tau }]{c_{\alpha
}^{(1)}}^{2}(k_{f}r))(f_{\alpha }^{(1)})^{2}\right ]
+\sum_{k=2,3}\left[ {f_{\alpha }^{(k)^{^{\prime }}}}^{2}{a_{\alpha
}^{(k)}}^{2}(k_{f}r)\right.\right. \nonumber \\&&\left. \left.
+\frac{2m}{\hbar^{2}}( \{V_{c}+V_{\sigma }+V_{\tau } +V_{\sigma \tau
}+(-6k+14)(V_{t\tau}+V_{t})-(k-1)(V_{ls\tau }+V_{ls})\right.\right.
\nonumber
\\&&\left.\left. +2[V_{T}+V_{\sigma T }+(-6k+14)V_{tT}-V_{\tau
z}]\}{a_{\alpha }^{(k)}}^{2}(k_{f}r)\right.\right. \nonumber
\\&&\left.\left. +[V_{l2}+V_{l2\sigma } +V_{l2\tau }+V_{l2\sigma \tau
}]{c_{\alpha }^{(k)}}^{2}(k_{f}r)+[V_{ls2}+V_{ls2\tau }] {d_{\alpha
}^{(k)}}^{2}(k_{f}r)) {f_{\alpha }^{(k)}}^{2}\right ]\right.
\nonumber \\&&\left. +\frac{2m}{\hbar^{2}}\{V_{ls}+V_{ls\tau
}-2(V_{l2}+V_{l2\sigma }+V_{l2\sigma \tau } +V_{l2\tau })-3(V_{ls2}
+V_{ls2\tau })\}b_{\alpha }^{2}(k_{f}r)f_{\alpha }^{(2)}f_{\alpha
}^{(3)}\right. \nonumber \\&&\left. +\frac{1}{r^{2}}(f_{\alpha
}^{(2)} -f_{\alpha }^{(3)})^{2}b_{\alpha }^{2}(k_{f}r)\right\},
 \end{eqnarray}
where $\alpha=\{J,L,S,S_z\}$ and the coefficient  ${a_{\alpha
}^{(1)}}^{2}$, etc., are defined as
\begin{eqnarray}\label{a1}
     {a_{\alpha }^{(1)}}^{2}(x)=x^{2}I_{L,S_{z}}(x),
 \end{eqnarray}
\begin{eqnarray}
     {a_{\alpha }^{(2)}}^{2}(x)=x^{2}[\beta I_{J-1,S_{z}}(x)
     +\gamma I_{J+1,S_{z}}(x)],
 \end{eqnarray}
\begin{eqnarray}
           {a_{\alpha }^{(3)}}^{2}(x)=x^{2}[\gamma I_{J-1,S_{z}}(x)
           +\beta I_{J+1,S_{z}}(x)],
      \end{eqnarray}
\begin{eqnarray}
     b_{\alpha }^{(2)}(x)=x^{2}[\beta _{23}I_{J-1,S_{z}}(x)
     -\beta _{23}I_{J+1,S_{z}}(x)],
 \end{eqnarray}
\begin{eqnarray}
         {c_{\alpha }^{(1)}}^{2}(x)=x^{2}\nu _{1}I_{L,S_{z}}(x),
      \end{eqnarray}
\begin{eqnarray}
        {c_{\alpha }^{(2)}}^{2}(x)=x^{2}[\eta _{2}I_{J-1,S_{z}}(x)
        +\nu _{2}I_{J+1,S_{z}}(x)],
 \end{eqnarray}
\begin{eqnarray}
       {c_{\alpha }^{(3)}}^{2}(x)=x^{2}[\eta _{3}I_{J-1,S_{z}}(x)
       +\nu _{3}I_{J+1,S_{z}}(x)],
 \end{eqnarray}
\begin{eqnarray}
     {d_{\alpha }^{(2)}}^{2}(x)=x^{2}[\xi _{2}I_{J-1,S_{z}}(x)
     +\lambda _{2}I_{J+1,S_{z}}(x)],
 \end{eqnarray}
\begin{eqnarray}\label{d2}
     {d_{\alpha }^{(3)}}^{2}(x)=x^{2}[\xi _{3}I_{J-1,S_{z}}(x)
     +\lambda _{3}I_{J+1,S_{z}}(x)],
 \end{eqnarray}
with
\begin{eqnarray}
          \beta =\frac{J+1}{2J+1},\ \gamma =\frac{J}{2J+1},\
          \beta _{23}=\frac{2J(J+1)}{2J+1},
 \end{eqnarray}
\begin{eqnarray}
       \nu _{1}=L(L+1),\ \nu _{2}=\frac{J^{2}(J+1)}{2J+1},\
       \nu _{3}=\frac{J^{3}+2J^{2}+3J+2}{2J+1},
      \end{eqnarray}
\begin{eqnarray}
     \eta _{2}=\frac{J(J^{2}+2J+1)}{2J+1},\ \eta _{3}=
     \frac{J(J^{2}+J+2)}{2J+1},
 \end{eqnarray}
\begin{eqnarray}
     \xi _{2}=\frac{J^{3}+2J^{2}+2J+1}{2J+1},\
     \xi _{3}=\frac{J(J^{2}+J+4)}{2J+1},
 \end{eqnarray}
\begin{eqnarray}
     \lambda _{2}=\frac{J(J^{2}+J+1)}{2J+1},\
     \lambda _{3}=\frac{J^{3}+2J^{2}+5J+4}{2J+1},
 \end{eqnarray}
and
\begin{eqnarray}
       I_{J,S_{z}}(x)=\int dq\ q^2 P_{S_{z}}(q) J_{J}^{2}(xq)\cdot
 \end{eqnarray}
In the last equation $J_{J}(x)$ is the Bessel function and
$P_{S_{z}}(q)$ is defined as
\begin{eqnarray}
P_{S_{z}}(q)&=&\frac{2}{3}\pi[(k_{F} ^{\sigma_{z1}})^{3}+(k_{F}
^{\sigma_{z2}})^{3}-\frac{3}{2}((k_{F} ^{\sigma_{z1}})^{2}+(k_{F}
^{\sigma_{z2}})^{2})q\nonumber\\ &-&\frac{3}{16}((k_{F}
^{\sigma_{z1}})^{2}-(k_{F} ^{\sigma_{z2}})^{2})^{2}q^{-1}+q^{3}]
 \end{eqnarray}
for                 $\frac{1}{2}|k_{F} ^{\sigma_{z1}}-k_{F}
^{\sigma_{z2}}|<q<\frac{1}{2}|k_{F} ^{\sigma_{z1}}+k_{F}
 ^{\sigma_{z2}}|$,
\begin{eqnarray}
P_{S_{z}}(q)=\frac{4}{3}\pi min((k_{F} ^{\sigma_{z1}})^3,(k_{F}
^{\sigma_{z2}})^3)
 \end{eqnarray}
for  $q<\frac{1}{2}|k_{F} ^{\sigma_{z1}}-k_{F}
 ^{\sigma_{z2}}|$, and
 \begin{eqnarray}
       P_{S_{z}}(q)=0
 \end{eqnarray}
for  $q>\frac{1}{2}|k_{F} ^{\sigma_{z1}}+k_{F}
 ^{\sigma_{z2}}|$, where $\sigma_{z1}$ or
 $\sigma_{z2}= +1,-1$ for spin up and down,
 respectively.

\subsection{Energy calculation of spin polarized neutron matter in the
presence of magnetic field}
Now we consider the case in which the spin polarized neutron
matter is under the influence of a strong magnetic field. Taking
the uniform magnetic field along the $z$ direction,
$B=B\widehat{k}$, the spin up and down particles correspond to
parallel and antiparallel spins with respect to the magnetic
field. Therefore, the contribution of magnetic energy of the
neutron matter is
\begin{eqnarray}
        E_{M}=-M_{z}B,
 \end{eqnarray}
where $M_{z}$ is the magnetization of the neutron matter which is
given by
\begin{eqnarray}
        M_{z}=N\mu_n\delta.
 \end{eqnarray}
In the above equation, $\mu_n= -1.9130427(5)$ is the neutron
magnetic moment (in units of the nuclear magneton). Consequently,
the energy per particle up to the two-body term in the presence of
magnetic field can be written as
 \begin{eqnarray}
 E([f])=E_{1}^{(B=0)}+E_{2}^{(B=0)}-\mu_n B \delta,
\end{eqnarray}
where $E_{1}^{(B=0)}$ and $E_{2}^{(B=0)}$ are given by Eqs.
(\ref{oneterm}) and (\ref{ener2}), respectively.
It should be noted that in usual thermodynamic treatments the
external magnetic field energy ($\frac{1}{8\pi}\int dV\ B^2$) is
usually left out since it does not affect the thermodynamic
properties of matter \cite{callen}. In fact the magnetic field
energy arises only from the magnetostatic energy in the absence of
matter,
but we are interested in the contribution of internal energy which
excludes the energy of magnetic field.
Therefore, the magnetic field contribution, $E_{mag}=\frac{B^2}{8
\pi }$, which is the \emph{energy density} (or ``magnetic
pressure'') of the magnetic field in the absence of matter is
usually omitted \cite{callen,Isayev}.

Now, we minimize the two-body energy with respect to the
variations in the function $f_{\alpha}^{(i)}$ subject to the
normalization constraint \cite{Bordbar57},
\begin{eqnarray}
        \frac{1}{N}\sum_{ij}\langle ij\left| h_{S_{z}}^{2}
        -f^{2}(12)\right| ij\rangle _{a}=0,
 \end{eqnarray}
where in the case of spin polarized neutron matter, the function
$h_{S_{z}}(r)$ is defined as follows,
\begin{eqnarray}
h_{S_{z}}(r)&=& \left\{\begin{array}{ll} \left[ 1-9\left(
\frac{J_{J}^{2}
(k_{F}^{(i)}r)}{k_{F}^{(i)}r}\right) ^{2}\right] ^{-1/2} &;~~ S_{z}=\pm1   \\ \\
1 &;~~ S_{z}= 0
\end{array}
\right.
\end{eqnarray}
From minimization of the two-body cluster energy, we get a set of
coupled and uncoupled differential equations which are the same as
those presented in Ref. \cite{Bordbar57}, with the coefficients
replaced by those indicated in Eqs. (\ref{a1})$-$(\ref{d2}). By
solving these differential equations, we can obtain correlation
functions to compute the two-body energy.

\section{RESULTS and DISCUSSION}\label{NLmatchingFFtex}
\begin{figure}

\includegraphics{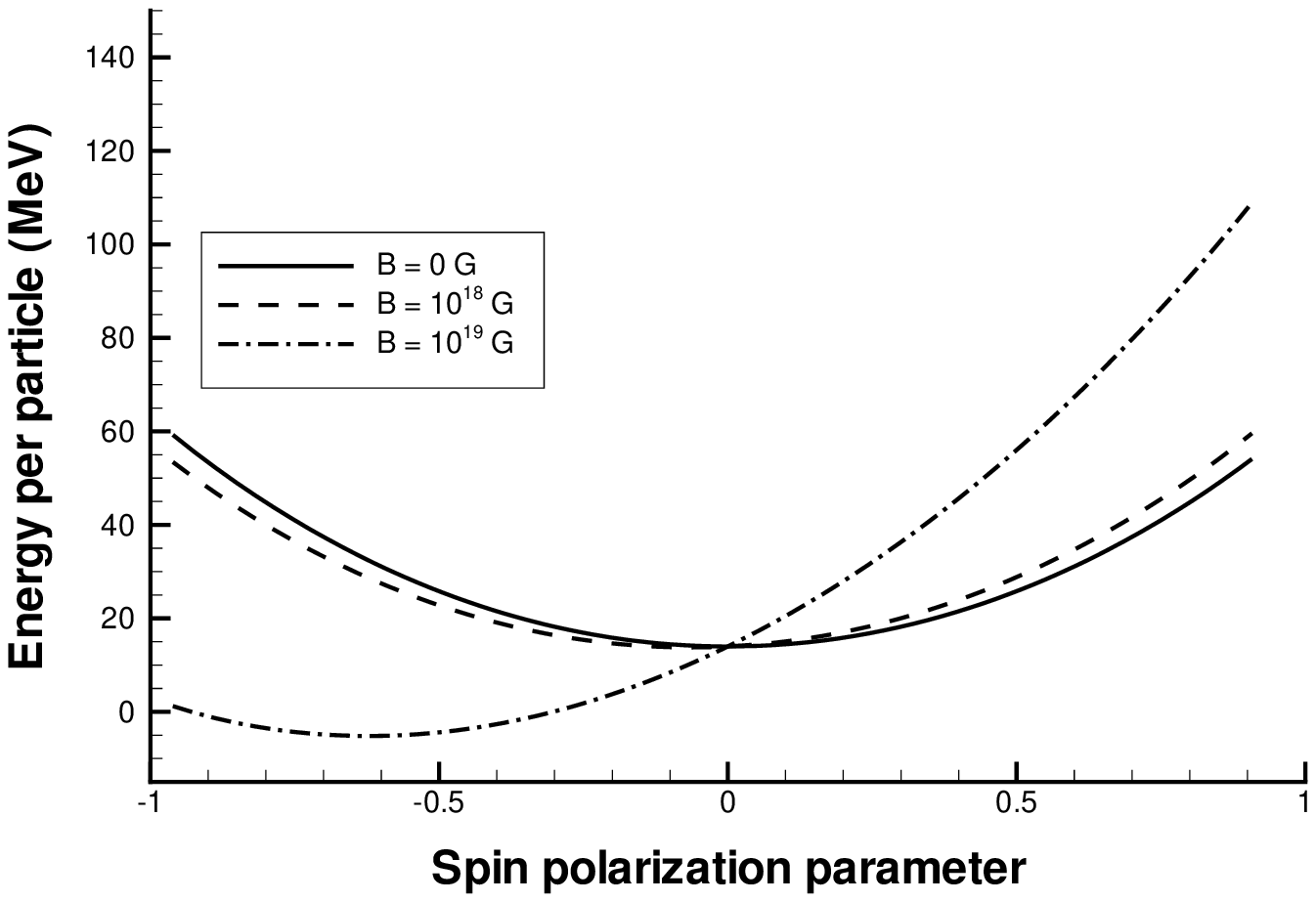}
 \caption{\label{fig:lj}The energy per particle versus the spin
 polarization parameter $(\delta)$ for different values of the magnetic field ($B$)
  at $\rho=0.2\ fm^{-3}$.}

\end{figure}

\begin{figure}

\includegraphics{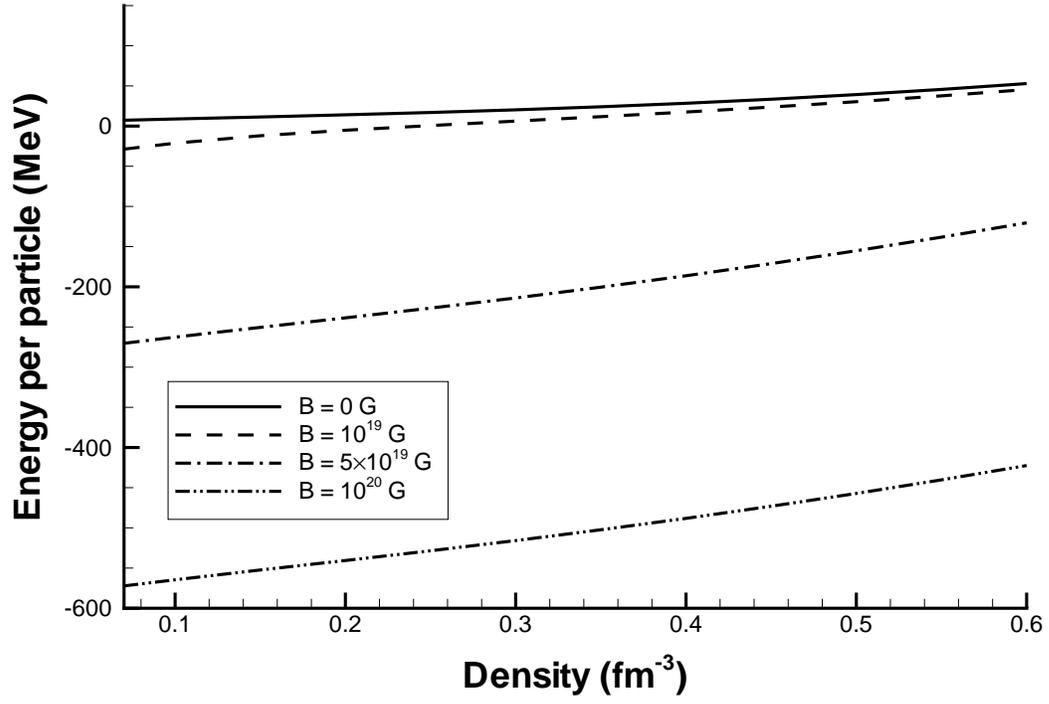}

\caption{\label{fig:2j}The ground state energy per particle as a
function of the density at different values of the magnetic field
($B$).}

\end{figure}

\begin{figure}

\includegraphics{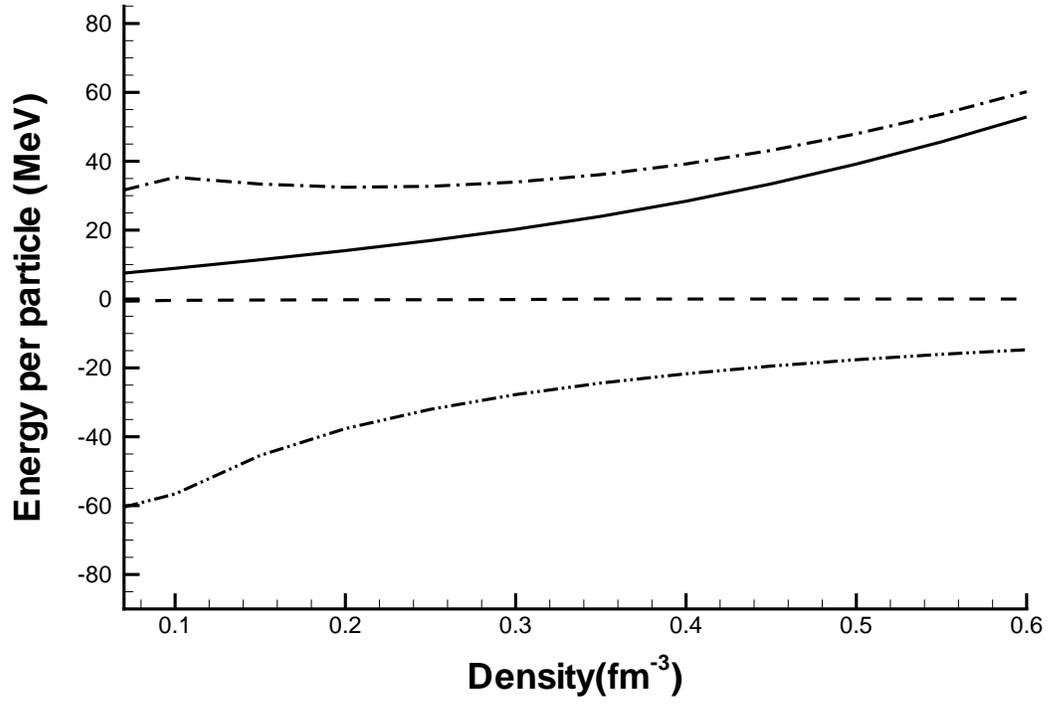}

\caption{\label{fig:12j}The energy contribution of spin polarized
neutron matter in the cluster expansion up to the two body term
($E_1 + E_2$) for the magnetic fields $B=10^{18}\ G$ (solid curve)
and $B=10^{19}\ G$ (dashed dotted curve), and the contribution of
magnetic energy ($E_M$) for magnetic fields $B=10^{18}\ G$ (dashed
curve) and $B=10^{19}\ G$ (dashed dotted dotted curve).}

\end{figure}

\begin{figure}

\includegraphics{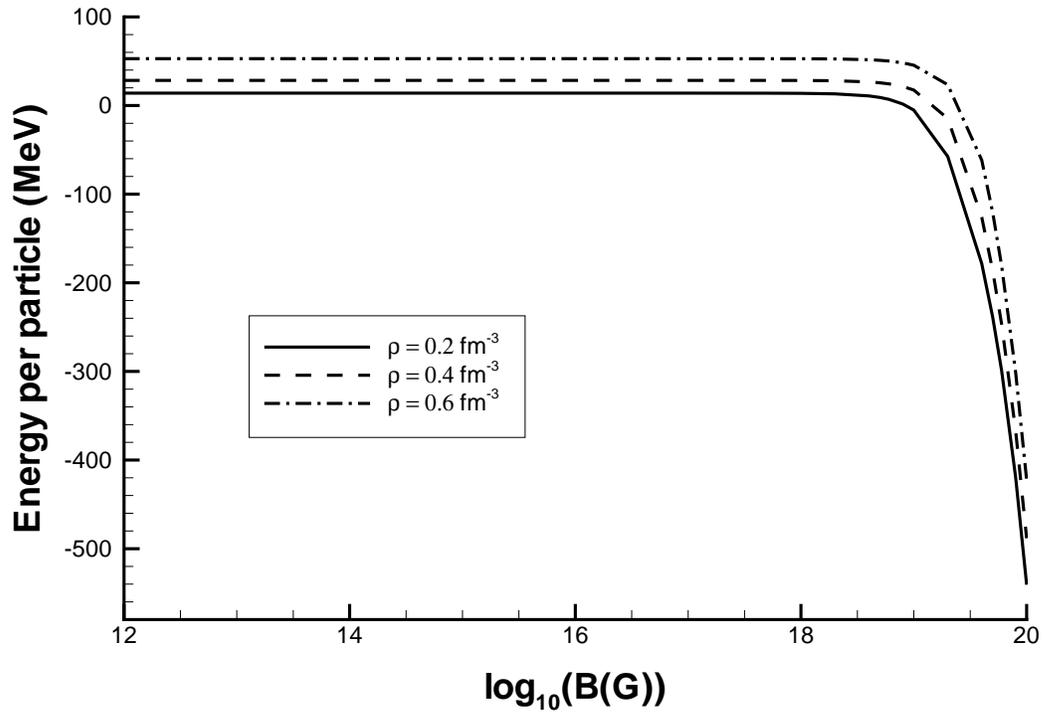}

\caption{\label{fig:7j}The ground state energy per particle as a
function of the magnetic field ($B$) at different values of the
density ($\rho$).}

\end{figure}

Our results for the energy per particle of spin polarized neutron
matter versus the spin polarization parameter for different values
of the magnetic field at $\rho=0.2\ fm^{-3}$ have been shown in Fig.
\ref{fig:lj}. We have found that for the values of magnetic field
below $10^{18}\ G$, the corresponding energies of different magnetic
fields are nearly identical. This shows that the effect of magnetic
field below $B\sim10^{18}\ G$ is nearly insignificant. From Fig.
\ref{fig:lj}, we can see that the spin polarization symmetry is
broken when the magnetic field is present and a minimum appears at
$-1< \delta <0$. By increasing the magnetic field strength from
$B\sim10^{18}\ G$ to $B\sim10^{19}\ G$, the value of spin
polarization corresponding to the minimum point approaches $-1$. We
also see that by increasing the magnetic field, the energy per
particle at minimum point(ground state energy) decreases, leading to
a more stable system.
For each density, we have found that above a certain value of the
magnetic field, the system reaches a saturation point and the
minimum energy occurs at $\delta=-1$.
For example at $\rho=0.2\ fm^{-3}$, for $B\gtrsim 1.8\times 10^{19}\
G$, the minimum energy occurs at $\delta=-1$. However, this
threshold value of the magnetic field increases by increasing the
density.
In Fig. \ref{fig:2j}, we have presented the ground state energy per
particle of spin polarized neutron matter as a function of the
density for different values of the magnetic field. For each value
of the magnetic field, it is shown that the energy per particle
increases monotonically by increasing the density. However, the
increasing rate of  energy versus  density increases by increasing
the magnetic field.
This indicates that at higher magnetic fields, the increasing rate
of the contribution of magnetic energy versus density is more than
that at lower magnetic fields.
In order to clarify this behavior, we have presented the energy
contribution of spin polarized neutron matter up to the two-body
term in the cluster expansion ($E_{1}+E_{2}$), and the magnetic
energy contribution ($E_M$) separately, as a function of density in
Fig. \ref{fig:12j}.
This figure shows that for the spin polarized neutron matter, the
difference between the magnetic energy contributions ($E_M$) of
different magnetic fields is substantially larger than that for the
energy contribution ($E_{1}+E_{2}$).
Fig. \ref{fig:7j} shows the ground state energy per particle of spin
polarized neutron matter as a function the magnetic field for
different values of density. We can see that by increasing the
magnetic field up to a value about $10^{18}\ G$, the energy per
particle slowly decreases, and then it rapidly decreases for the
magnetic fields greater than this value. This indicates that above
$B\sim10^{18}\ G$, the effect of magnetic field on the energy
construction of the spin polarized neutron matter becomes more
important.

\begin{figure}

\includegraphics{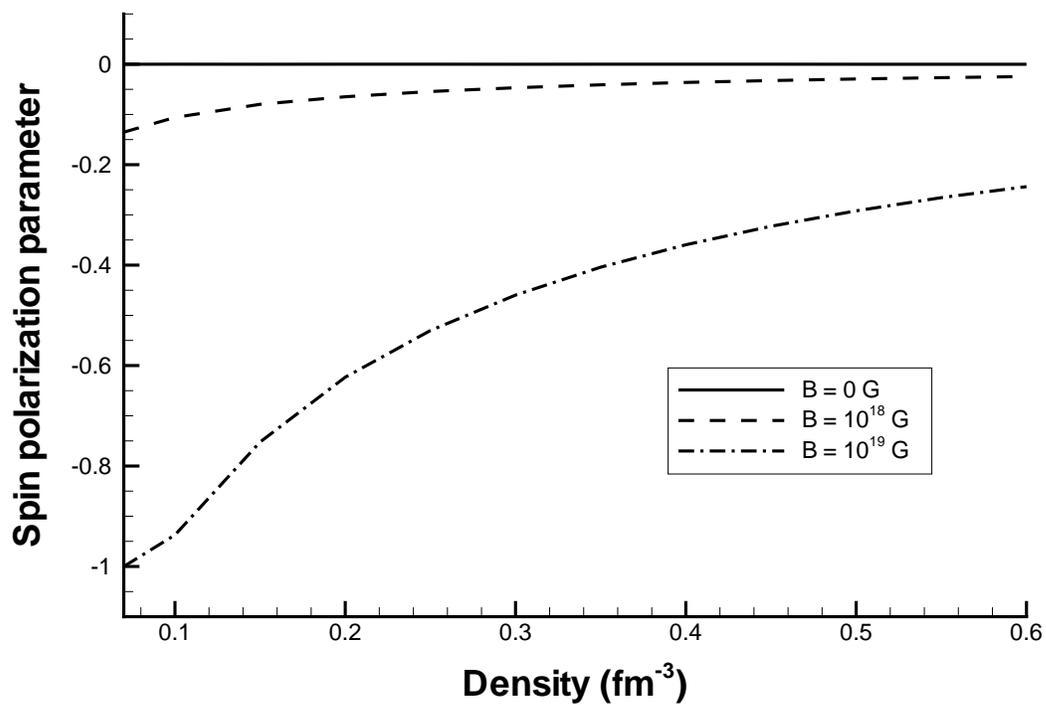}

 \caption{\label{fig:3j}The spin polarization parameter at the
 equilibrium state of the system as a function of the density at
 different values of the magnetic field ($B$).}
\end{figure}

\begin{figure}

\includegraphics{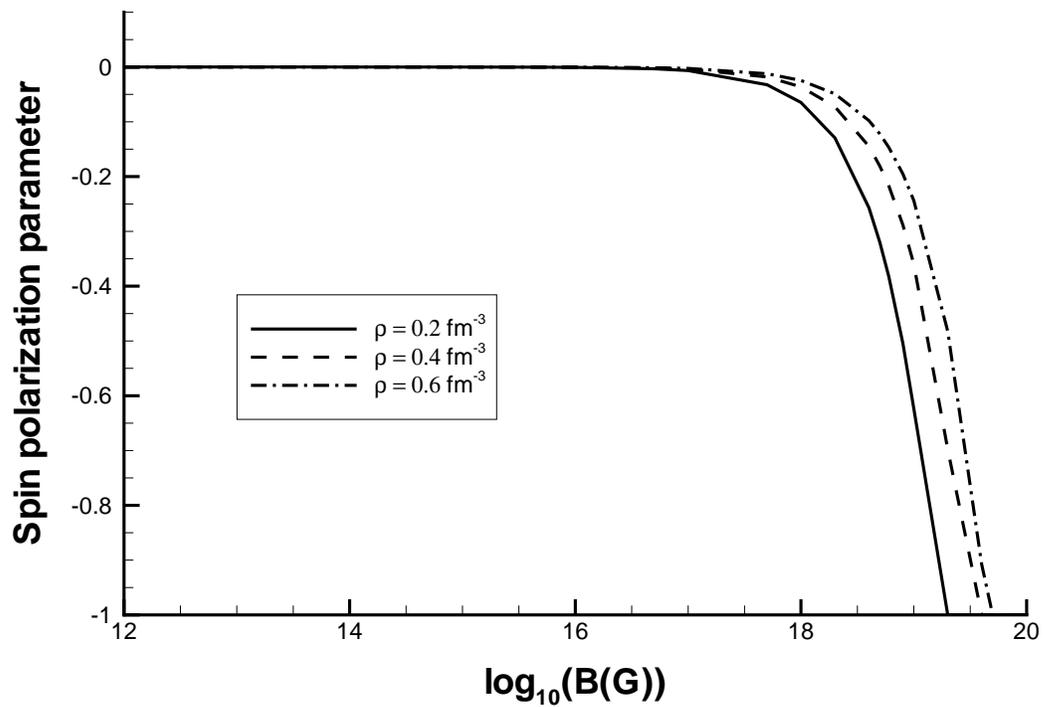}

 \caption{\label{fig:8j}The spin polarization parameter corresponding to the
equilibrium state of the system as a function of the magnetic field
($B$) at different values of the density ($\rho$).}
\end{figure}

In Fig. \ref{fig:3j}, the spin polarization parameter corresponding
to the equilibrium state of the system is plotted as a function of
density for different values of the magnetic field.
It is seen that at each magnetic field, the magnitude of spin
polarization parameter decreases by increasing the density.
Fig. \ref{fig:3j} also shows that for the magnetic fields below
$10^{18}\ G$, at high densities, the system nearly becomes
unpolarized. However, for higher magnetic fields, the system has a
substantial spin polarization, even at high densities.
In Fig. \ref{fig:8j}, we have plotted the spin polarization
parameter at the equilibrium as a function of the magnetic field at
different values of  density.
This figure shows that below $B\sim10^{18}\ G$, no anomaly is
observed and the neutron matter can only be partially polarized.
This partial polarization is maximized at lower densities and
amounts to about $14\%$ of its maximum possible value of $-1$ . From
Fig. \ref{fig:8j}, we can also see that below $B\sim10^{17}\ G$, the
spin polarization parameter is nearly zero. This clearly confirms
the absence of the magnetic ordering for the neutron matter up to
$B\sim10^{17}\ G$.
For the magnetic fields greater than about $10^{18}\ G$, it is
shown that the magnitude of spin polarization rapidly increases by
increasing the magnetic field. This shows a ferromagnetic phase
transition in the presence of a strong magnetic field.
For each density, we can see that at high magnetic fields, the value
of spin polarization parameter is close to $-1$. The corresponding
value of the magnetic field increases by increasing the density.

\begin{figure}

\includegraphics{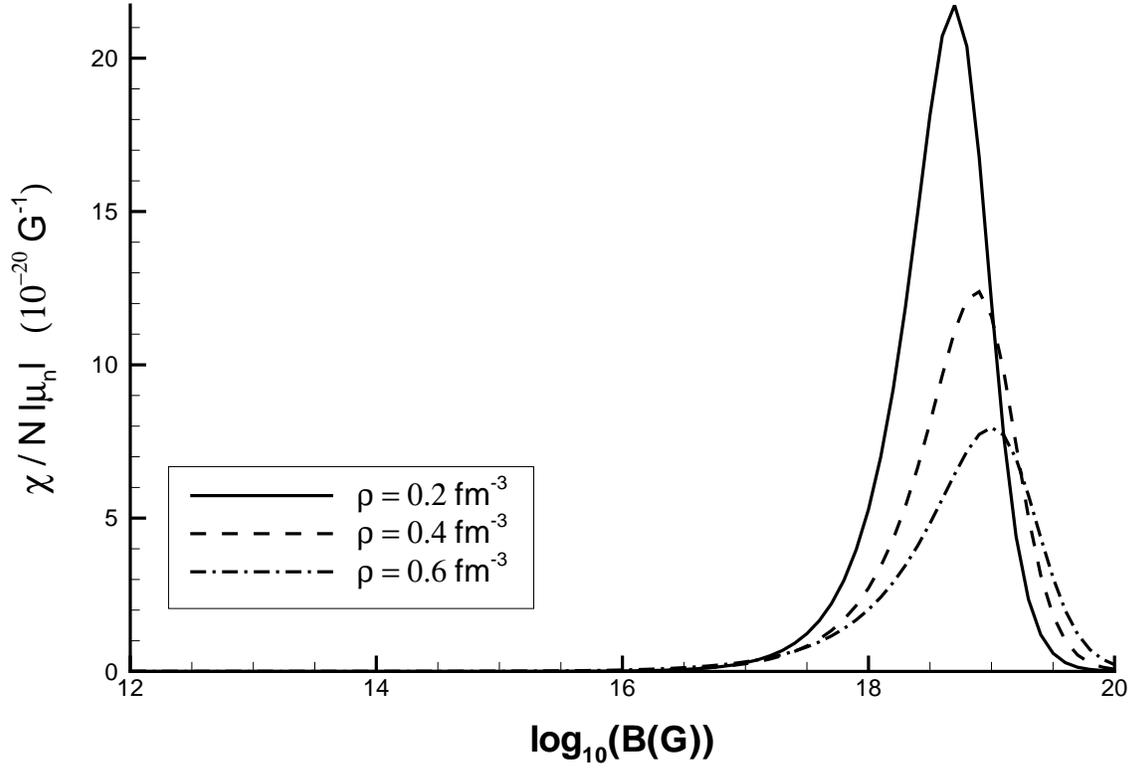}

 \caption{\label{fig:6j}
The magnetic susceptibility ($\chi/N|\mu_{n}|$) as a function of the
magnetic field ($B$) at different values of the density ($\rho$).}
\end{figure}

\begin{figure}

\includegraphics{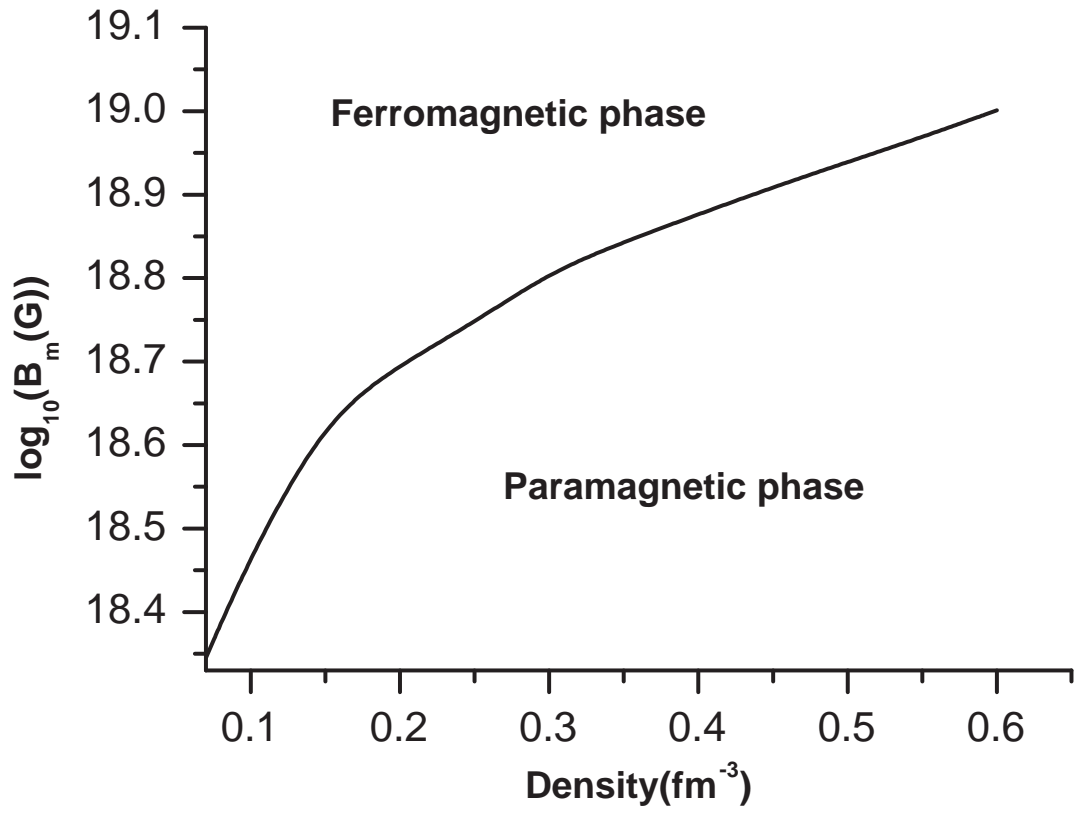}

 \caption{\label{fig:4j}
Phase diagram for the spin polarized neutron matter in the
presence of strong magnetic field.}
\end{figure}

The magnetic susceptibility ($\chi$) which characterizes the
response of a system to the magnetic field, is defined by
\begin{eqnarray}
     \chi(\rho,B)={\left(\frac{\partial M_{z}(\rho,B)}{\partial B} \right)
_{\rho}}
 \end{eqnarray}
In Fig. \ref{fig:6j}, we have plotted the ratio $\chi/N|\mu_{n}|$
for the spin polarized neutron matter versus the magnetic field at
three different values of the density. As can be seen from Fig.
\ref{fig:6j},  for each density, this ratio shows a maximum at a
specific magnetic field. This result confirms the existence of the
ferromagnetic phase transition induced by the magnetic field. We
see that the magnetic field at phase transition point, $B_{m}$,
depends on the density of the system.
Fig. \ref{fig:4j} shows the phase diagram for the spin polarized
neutron matter. We can see that by increasing the density, $B_{m}$
grows monotonically. It explicitly means that at higher densities,
the phase transition occurs at higher values of the magnetic field.

From the energy of spin polarized neutron matter, at each magnetic
field, we can  evaluate the corresponding pressure ($P_{kinetic}$)
using the following relation,

\begin{eqnarray}
      P_{kinetic}(\rho,B)= \rho^{2}
      {\left(\frac{\partial E(\rho,B)}
      {\partial \rho} \right)
_{B}}
 \end{eqnarray}
Our results for the kinetic pressure of spin polarized neutron
matter versus the density for different values of the magnetic field
have been shown in Fig. \ref{fig:5j}.  It is obvious that with
increasing the density, the difference between the pressure of spin
polarized neutron matter at different magnetic field becomes more
appreciable.
Fig. \ref{fig:5j} shows that the equation of state of the spin
polarized neutron matter becomes stiffer as the magnetic field
strength increases. This stiffening is due to the inclusion of
neutron anomalous magnetic moments.
This is in agreement with the results obtained in Refs.
\cite{Brod0,Yue6}.
It should be noted here that to find the total pressure related for
the neutron star structure, the contribution from the magnetic
field, $P_{mag}=\frac{B^2}{8 \pi}$, should be added to the kinetic
pressure \cite{Brod0,Brod2}.
However, in this work we are not interested in the neutron star
structure and have thus omitted the contribution of ``magnetic
pressure'' in our calculations for neutron matter \cite{Isayev}.
This term, if included, simply adds a constant amount to the curves
depicted in Fig. \ref{fig:5j}.

\section{Summary and Concluding Remarks}
We have recently calculated several properties of the spin polarized
neutron matter in the absence of magnetic field using the lowest
order constrained variational method with $AV_{18}$ potential.
In this work, we have generalized our calculations for spin
polarized neutron matter in the presence of strong magnetic field at
zero temperature using this method. We have found that the effect of
magnetic fields below $B\sim10^{18}\ G$ is almost negligible. It was
shown that in the presence of magnetic field, the spin polarization
symmetry is broken and the energy per particle shows a minimum at
$-1< \delta <0$, depending on the strength of the magnetic field. We
have shown that the ground state energy per particle decreases by
increasing the magnetic field. This leads to a more stable system.
It is seen that the increasing rate of energy versus  density
increases by increasing the magnetic field. Our calculations show
that above $B\sim10^{18}\ G$, the effect of magnetic field on the
properties of neutron matter becomes more important. In the study of
spin polarization parameter, we have shown that for a fixed magnetic
field, the magnitude of spin polarization parameter at the minimum
point of energy decreases with increasing  density. At strong
magnetic fields with strengths greater than $10^{18}\ G$, our
results show that a field-induced ferromagnetic phase transition
occurs for the neutron matter. By investigating the magnetic
susceptibility of the spin polarized neutron matter, it is clear
that as  the density increases, the phase transition occurs at
higher values of the magnetic field. Through the calculation of
 pressure as a function of density at different values of the
magnetic field, we observed the stiffening of the equation of
state in the presence of the magnetic field.

\begin{figure}

\includegraphics{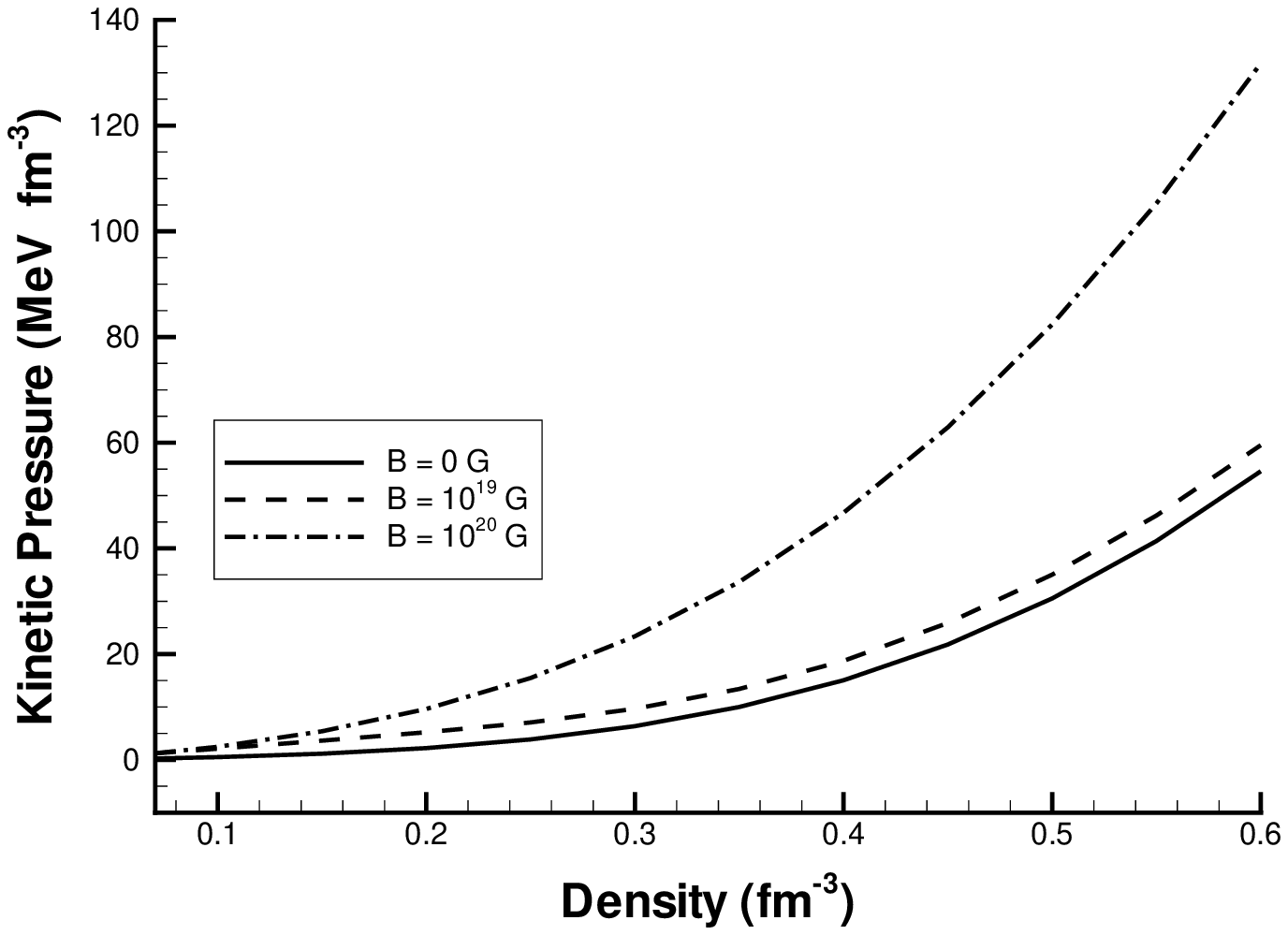}

 \caption{\label{fig:5j}The equation of state of spin polarized neutron matter for
  different values of the magnetic field ($B$).}

\end{figure}

Finally, we would like to address the question of thermodynamic
stability of such neutron stars at ultra-high magnetic fields. One
may wonder if the effect of magnetic pressure,
$P_{mag}=\frac{B^2}{8 \pi}$, which we have omitted here, is added
to the kinetic pressure $P_{kinetic}$, then at ultra-strong
magnetic fields, the system might become gravitationally unstable
due to excessive outward pressure. For the fields considered in
this work (up to $10^{20}\ G$), this scenario does not seem likely
\cite{shap}. We note that the increase of magnetic field leads to
stiffening of the equation of state (Fig. \ref{fig:5j}) which in
turn leads to larger mass and radius for the neutron star
\cite{bordbarnew}. This in turn increases the effect of
gravitational energy, offsetting the increased pressure. We also
note that the existence of a well-defined thermodynamic energy
minimum for all fields considered in our work indicates the
thermodynamic stability of our system. The existence of such
well-defined minimum energy is unaffected by the addition of
magnetic energy. The detailed analysis of such situations along
with accompanying change in proton fraction is a possible avenue
for future research.

\acknowledgements{We would like to thank two anonymous referees for
constructive criticisms. This work has been supported by Research
Institute for Astronomy and Astrophysics of Maragha. We wish to
thank Shiraz University Research Council.}


\newpage
\newpage
\newpage
\newpage
\newpage

\newpage
\newpage
\newpage
\newpage


\begin{thebibliography}{99}

\bibitem{Reisen}  A. Reisenegger, {Astron. Nachr.} {\bf 328}, 1173 (2007).
\bibitem{Woltjer} L. Woltjer, {Astrophys. J.} {\bf 140}, 1309 (1964).
\bibitem{Tayler}  R.J. Tayler, {MNRAS} {\bf 161}, 365 (1973).
\bibitem{Spruit}  H. Spruit, {Astron. Astrophys.} {\bf 381}, 923 (2002).
\bibitem{Thompson} C. Thompson and R. C. Duncan, {Astrophys. J.} {\bf 408}, 194 (1993).
\bibitem{Lai} D. Lai and S. L. Shapiro, {Astrophys. J.} {\bf 383}, 745 (1991).
\bibitem{shap} S. Shapiro and S. Teukolsky, \emph{Black Holes, White Dwarfs
and Neutron Stars}, (Wiley-New York, 1983).
\bibitem{Yuan} Y. F. Yuan  and J. L. Zhang , {Astron. Astrophys.} {\bf 335}, 969 (1998).
\bibitem{zhang} Y. F. Yuan and J. L. Zhang, {Astrophys. J.} {\bf
525},
950 (1999).
\bibitem{Brod0} A. Broderick,  M. Prakash and J. M. Lattimer, {Astrophys. J.} {\bf
537},
351 (2000).
\bibitem{SUH}  I.S. Suh and G. J. Mathews, {Astrophys. J.} {\bf
546},
1126 (2001).
\bibitem{Chen} W. Chen, P. Q. Zhang and L. G. Liu, {Mod. Phys. Lett.}  {\bf A
22},
623 (2007).
\bibitem{Yue6} P. Yue and H. Shen, {Phys. Rev.}  {\bf C 74},
045807 (2006).
\bibitem{Brod2} A. Broderick, M. Prakash, and J. M. Lattimer, {Phys. Lett}. {\bf B
531},
167 (2002).
\bibitem{Chakra7} S. Chakrabarty, D. Bandyopadhyay, and S. Pal, {Phys. Rev.
Lett.}  {\bf 78}, 2898 (1997).
\bibitem{Isayev}  A. A. Isayev and J. Yang, {Phys. Rev.}  {\bf C 80}, 065801 (2009).
\bibitem{Isayev1}  A. A. Isayev and J. Yang, {J. Korean Astronom. Soc}. {\bf 43}, 161 (2010).
\bibitem{Garcia8} M. A. Perez-Garcia, {Phys. Rev.} {\bf C 77},
065806 (2008).
\bibitem{Garcia9} M. A. Perez-Garcia, {Phys. Rev.} {\bf C 80},
045804 (2009).
\bibitem{Garcia10} M. A. Perez-Garcia, J. Navarro, and A. Polls, {Phys. Rev.} {\bf C
80},
025802 (2009).
\bibitem{ANAND}  J. D. Anand, N. Chandrika Devi, V. K. Gupta, and S.
Singh, {Astrophys. J.} {\bf 538}, 870 (2000).
\bibitem{Ghosh} S. Ghosh and S. Chakrabarty, {Pramana} {\bf 60}, 901 (2002).
\bibitem{Chakra6} S. Chakrabarty, {Phys. Rev}. {\bf D 54},
1306 (1996).
\bibitem{Gupta}  V.K.Gupta, A. Gupta, S.Singh and J.D.Anand,
{Int. J. Mod. Phys}. {\bf D 11}, 545 (2002).
\bibitem{Bandyo} D. Bandyopadhyay, S. Chakrabarty and S. Pal,
{Phys. Rev. Lett}. {\bf 79}, 2176 (1997).
\bibitem{bord-pey} G. H. Bordbar and A. Peyvand (2010) submitted
for publication.
\bibitem{Cardall} C.Y. Cardall, M. Prakash and J.M. Lattimer,
 {Astrophys. J.} {\bf 554}, 322 (2001).
\bibitem{Isayev2}  A. A. Isayev, {Phys. Rev}. {\bf C 74}, 057301 (2006).
\bibitem{Bordbar75} G. H. Bordbar and M. Bigdeli, {Phys. Rev}. {\bf C
75},
045804 (2007).
\bibitem{Bordbar76} G. H. Bordbar and M. Bigdeli, {Phys. Rev}. {\bf C
76},
035803 (2007).
\bibitem{Bordbar77} G. H. Bordbar and M. Bigdeli, {Phys. Rev}. {\bf C
77},
015805 (2008).
\bibitem{Bordbar78} G. H. Bordbar and M. Bigdeli, {Phys. Rev}. {\bf C
78},
054315 (2008).
\bibitem{Bordbar80} M. Bigdeli, G. H. Bordbar  and Z. Rezaei, {Phys. Rev}. {\bf C
80},
034310 (2009).
\bibitem{Bordbar81} M. Bigdeli, G. H. Bordbar  and A. Poostforush, {Phys. Rev}.
{\bf C 82}, 034309 (2010).

\bibitem{Jastrow}J. W. Clark, Prog. Part. Nucl. Phys. {\bf 2}, 89 (1979).
\bibitem{Wiringa} R. B. Wiringa, V. Stoks, and R. Schiavilla, {Phys. Rev}. {\bf C 51}, 38 (1995).
\bibitem{Owen} J. C. Owen, R. F. Bishop, and J. M. Irvine, {Nucl. Phys}. {\bf A 277}, 45 (1977).
\bibitem{callen} H. B. Callen, \emph{Thermodynamics and
an Introduction to Thermostatistics}, (John Wiley $\&$ Sons, Inc,
1985).
\bibitem{Bordbar57} G. H. Bordbar and M. Modarres, {Phys. Rev}. {\bf C 57}, 714 (1998).
\bibitem{bordbarnew} G. H. Bordbar and M. Hayati, Int. J. Mod. Phys.
{\bf A 21}, 1555 (2006).
\end{thebibliography}
\end{document}